\documentclass[11pt]{article}

\newcommand{\I}[1]{\mathcal{I}_{#1}}
\newcommand{\J}[1]{\mathcal{H}_{#1}}
\newcommand{\JJ}[1]{\mathsf{H}_{#1}}
\newcommand{\VV}{\mathsf{V}}
\newcommand{\TT}{\mathsf{T}}
\renewcommand{\SS}{\mathsf{S}}
\newcommand{\FF}{\mathsf{F}}
\newcommand{\fJ}[1]{\widetilde{\mathcal{H}}_{#1}}
\newcommand{\F}[1]{\mathcal{F}_{#1}}
\newcommand{\alp}{\alpha_{\scriptscriptstyle +}}
\newcommand{\alm}{\alpha_{\scriptscriptstyle -}}
\newcommand{\qq}{\mathbf{q}}
\newcommand{\scsc}{{\scriptscriptstyle}}
\newcommand{{\zo}}{\ensuremath{z_{\scsc 0}}}
\newcommand{\loga}[1]{\log\!\left(#1\right)}

\newcommand{\Sp}[1]{\mathrm{Li}_{2}\!\left(#1\right)}
\newcommand{\sgn}{\mathrm{sgn}}

\newcommand{\oo}{{\cal O}}
\newcommand{\mub}{\overline{\mu}}
\date{December 10, 2001}
\title{One-loop vertex integrals in heavy-particle effective theories}
\author{Antonio O.\ Bouzas \thanks{E-mail:
    abouzas@mda.cinvestav.mx}\\\small Departamento de F\'{\i}sica
  Aplicada, CINVESTAV-IPN \\\small Apdo.\ Postal 73 ``Cordemex,''
  M\'erida 97310, Yucat\'an, M\'exico 
  \and 
  Rub\'en Flores-Mendieta\\\small Instituto de F{\'\i}sica,
  Universidad Aut\'onoma de San Luis Potos{\'\i}\\\small  
  \'Alvaro Obreg\'on 64, Zona Centro, San Luis Potos{\'\i}, S.L.P.\
  78000, M\'exico}
\begin{document}
\maketitle
\begin{abstract}
  We give a complete analytical computation of three-point one-loop
  integrals with one heavy propagator, up to the third tensor rank,
  for arbitrary values of external momenta and masses.
\end{abstract}
\section{Introduction}
\label{sec:intro}

The study of the dynamics and spectroscopy of hadrons containing a
heavy quark has been greatly simplified and systematized with the
introduction of heavy quark effective theory (HQET) \cite{hqet}.
Heavy-particle theories along similar lines have also been successfully
applied in other, related contexts.  Thus, in those cases where a
chiral approach to the strong interactions of heavy hadrons with light
mesons is applicable, a combination of chiral and heavy-quark
symmetries leads to heavy hadron chiral perturbation theory (HHChPT)
\cite{hqet}.  A heavy-particle expansion has also been developed in
the chiral-perturbative framework for nucleon-meson interactions,
which constitutes the so-called heavy baryon chiral perturbation
theory (HBChPT) \cite{hbchpt}.

In this paper we report on a complete analytic computation of
three-point loop integrals with one heavy propagator, up to the third
tensor rank, for arbitrary real values of masses and residual momenta.
Together with the results for vertex integrals with two heavy
propagators \cite{bouzas}, this gives the set of all one-loop
three-point integrals in heavy particle theories.

For the calculations we use the same method as in \cite{bouzas},
namely, we compute the vertex integrals with heavy propagators as
limits of ordinary three-point integrals.  The latter are evaluated by
closely following the method of \cite{thooft}, so that the calculation
is standardized and the results expressed in terms of the usual logs
and dilogs.  Since we only have to compute to leading order in a
certain limit, however, the calculations are drastically simplified as
compared to ordinary vertex integrals.

In the next section, we define the integrals to be studied and give
the calculational details and results for the scalar integral.  We
also discuss some particular cases which are of importance in
applications and as cross-checks of our results.  Then, in section
\ref{sec:tensorint} we consider the vector, second-, and third-rank
tensor integrals, which are given in terms of scalar integrals.

\section{Vertex integrals}
\label{sec:vertex}

The three-point loop integrals we consider are of the form, 
\begin{equation} 
  \J{3}^{\alpha_1\cdots \alpha_n}  =  \frac{i\mu^{4-d}}{(2\pi)^d}
  \int\! d^d \ell \frac{\ell^{\alpha_1}\cdots \ell^{\alpha_n}}{
    \left(2v\!\cdot\! \ell-\delta M+i\varepsilon\right)
    \left((\ell-p)^2-m^2+i\varepsilon\right)
    \left((\ell-p^\prime)^2-m^{\prime 2}+i\varepsilon\right)}. \label{j3}
\end{equation}
Here $v^\mu$ is the velocity of the heavy lines, $p^\mu$ and
$p^{\prime\mu}$ the residual momenta of the external heavy particles,
and $\delta M$ the residual mass of the internal one, given by the
mass splitting within the corresponding heavy-quark symmetry
multiplet. $m$ and $m^\prime$ are the masses of the light particles in
the loop, which are pseudoscalar mesons in HHChPT and HBChPT, and
(massless) gluons in HQET.  The scalar integral is finite, whereas the
vector and higher-rank tensors are divergent, with degree of
divergence $n+d-5$. We consider tensor integrals with $n=1,2,3$ below.
These are computed in terms of integrals with lower rank and fewer
points.  

By shifting the integration variable $\ell\rightarrow \ell + (p^\prime
+ p)/2$ , the scalar integral can be written as,
\begin{equation}\label{j3s}
    \J{3} =  \frac{i}{(2\pi)^4}
  \int\! d^4 \ell \frac{1}{
    \left(2v\!\cdot\! \ell-\Delta+i\varepsilon\right)
    \left((\ell+q)^2-m^2+i\varepsilon\right)
    \left((\ell-q)^2-m^{\prime 2}+i\varepsilon\right)}, 
\end{equation}
with $\Delta=\delta M - v\cdot (p^\prime + p)$, $q=(p^\prime - p)/2$.
Thus, $\J{3}$ depends explicitly on $q^\mu$, through $q^2$ and $v\cdot
q$, but depends on $(p^\prime + p)^\mu$ only through $\Delta$.  Since
$\J{3}$ is invariant under the transformation $q\rightarrow -q$, $m
\leftrightarrow m^\prime$ of its arguments, we can choose $v\cdot q >
0$ without loss of generality.

We will compute the scalar integral $\J{3}$ as the large-mass limit of
the auxiliary integral
\begin{equation}
  \label{eq:fj3s}
  \fJ{3}  = \frac{i}{(2\pi)^4}
  \int\! d^4 \ell \frac{1}{
    \left((\ell+M v)^2- (M+\Delta/2)^2+i\varepsilon\right)
    \left((\ell+q)^2- m^2+i\varepsilon\right)
    \left((\ell-q)^2-m^{\prime 2}+i\varepsilon\right)}.
\end{equation}
Since both $\J{3}$ and $\fJ{3}$ are convergent, we clearly have,
$  \J{3} = \lim_{M\rightarrow\infty} M \fJ{3}.$
The calculation of $\fJ{3}$ to leading order in $M$ will be carried
out by closely following the standard method of \cite{thooft}.  We
will only sketch the main steps of the calculation.

Starting from the usual Feynman parametrization for $\fJ{3}$, after
integrating over $d^4\ell$ and over the Feynman parameter multiplying
the last propagator in (\ref{eq:fj3s}), we obtain,
\begin{eqnarray}
  \label{eq:fj3a}
  \fJ{3} & = & \frac{1}{(4\pi)^2} \int_0^1\! dz\int_0^{1}\! dx\,
  \frac{1}{D} + \oo \left(\frac{1}{M^2}\right) \\
  D & = & M^2 z^2+M z (-2 (1-2x) v\cdot q+\Delta) + x m^2 + (1-x)
  m^{\prime 2} - 4 x (1-x) q^2 - i\varepsilon \nonumber.
\end{eqnarray}
In this expression we have retained only the leading terms in $M$ in
the coefficients of the polynomial in the denominator.  Clearly, only
the region where $z = \oo (M^{-1})$ will contribute to $\fJ{3}$ to
$\oo (M^{-1})$, so we have also set the upper limit of the integral
over $x$ to 1 instead of $1-z$.

We notice that if we make the change of variable $\lambda=Mz$ in
(\ref{eq:fj3a}) and let $M\rightarrow\infty$, we are led to the HQET
parametrization for $\J{3}$, confirming the validity of the
approximations leading to (\ref{eq:fj3a}).   

Following \cite{thooft}, we make the change of variable $z\rightarrow
z-\alpha x/M$, with $\alpha$ one of the roots to $(\alpha v-2 q)^2=0$, 
\begin{equation}
  \label{eq:alpha}
  \alpha_\pm = 2 (v\cdot q \pm |\mathbf{q}|).
\end{equation}
We henceforth assume that $q^2 < (v\cdot q)^2$, so that $|\qq|
\equiv \sqrt{(v\cdot q)^2 - q^2} > 0$ is real.  For the change of
variable we choose $\alpha=\alp >0$, leading to, 
\begin{equation}
  \label{eq:fj3b}
 \fJ{3}  =  \frac{1}{(4\pi)^2} \int_0^1\!
 dx\int_{\alpha x/M}^{1+\alpha x/M}\! dz\,
 \frac{1}{M^2 z^2-(4 M|\qq| z-\zo) x + M z (-2 v\cdot q + \Delta)
 + m^{\prime 2} -i \varepsilon},
\end{equation}
with $\zo = -(m^{\prime 2}-m^2) - \alp (\Delta-2 |\qq|)$.
Exchanging the order of integration, 
\begin{eqnarray}
  \label{eq:fj3c}
 \fJ{3}  & = & \frac{1}{(4\pi)^2} \left\{
 \int_0^{\alpha/M}\!\! dz \int_0^{Mz/\alpha}\!\! dx 
 +  \int_{\alpha/M}^{1}\!\! dz \int_0^1\!\! dx
 + \int^{1+\alpha/M}_{1}\!\! dz \int_{\frac{M}{\alpha}(z-1)}^1\!\! dx \right\}
\\
 & & \times  \frac{1}{M^2 z^2-(4 M|\qq| z-\zo) x + M z (-2 v\cdot q + \Delta)
   + m^{\prime 2} -i \varepsilon}\nonumber.
\end{eqnarray}
In the last integral in this equation $z = \oo(1)$ over the entire
domain of integration, so that the integral itself is $\oo (M^{-2})$
and can be neglected.  We will drop it henceforth. Integrating over
$dx$ we then have,
\begin{eqnarray}
  \label{eq:fj3d}
\fJ{3}  & = & \frac{1}{(4\pi)^2} \int_0^{\alpha/M} \!\! dz 
              \frac{-1}{4 M |\qq| z-\zo} \log\left( \frac{Y}{Z}\right) 
              +  
              \frac{1}{(4\pi)^2} \int_{\alpha/M}^1 \!\! dz 
              \frac{-1}{4 M |\qq| z-\zo} \log\left( \frac{X}{Z}\right)
              \\
X & = & M^2 z^2 + M z (-2 v\cdot q+\Delta) + m^{\prime 2} - i
   \varepsilon - (4 M |\qq| z-\zo) \nonumber \\
Y & = & M^2 z^2 + M z (-2 v\cdot q+\Delta) + m^{\prime 2} - i
   \varepsilon - \frac{M}{\alpha} z (4 M |\qq| z-\zo) \nonumber \\
Z & = & M^2 z^2 + M z (-2
   v\cdot q+\Delta) + m^{\prime 2} - i \varepsilon \nonumber
\end{eqnarray}
We notice that there is no singularity at the zero of the denominator
of the integrand in (\ref{eq:fj3d}),
$z = \zo/(4 M |\qq|)$, since $X = Y = Z$ there and the logs
vanish.  The roots of these polynomials in $z$ are given by,
\begin{eqnarray}
  \label{eq:roots}
  x_{1,2} & = & v\cdot p^\prime + 2 |\qq| -\frac{\delta
    M}{2} \pm \sqrt{\left(v\cdot p-\frac{\delta M}{2}\right)^2 - m^2 +
    i \varepsilon} \nonumber\\
  y_{1,2} & = & \frac{1}{2\alm} \left( 4q^2 + m^{\prime 2} - m^2 \pm
    \sqrt{\left(4 q^2 - (m^\prime + m)^2 \rule{0ex}{2ex}\right)
      \left(4 q^2 - (m^\prime - m)^2 \rule{0ex}{2ex}\right)
      + i \varepsilon\sigma}\right) \\
  z_{1,2} & = & v\cdot p^\prime -\frac{\delta M}{2} \pm
    \sqrt{\left(v\cdot p^\prime -\frac{\delta M}{2}\right)^2 -
    m^{\prime 2} + i \varepsilon}, \nonumber
\end{eqnarray}
where in the expression for $y_j$ we denoted $\sigma\equiv \sgn
(q^2)$.  The fact that $X$, $Y$, and $Z$ are all equal at $z = \zo/(4
M |\qq|)$ leads to the important identity,
\begin{equation}
  \label{eq:ident}
  \left(\frac{\zo}{4 |\qq|} - x_1\right) \left(\frac{\zo}{4 |\qq|} -
  x_2\right) = \frac{\alm}{\alp} 
  \left(\frac{\zo}{4 |\qq|} - y_1\right) \left(\frac{\zo}{4 |\qq|} -
  y_2\right) = 
  \left(\frac{\zo}{4 |\qq|} - z_1\right) \left(\frac{\zo}{4 |\qq|} -
  z_2\right)
\end{equation}

Factorizing the polynomials in the arguments of the logarithms in
(\ref{eq:fj3d}), we get
\begin{eqnarray}
  \label{eq:fj3e}
\fJ{3}  & = & \frac{1}{(4\pi)^2} \int_0^{\alpha/M} \!\! dz 
\frac{-1}{4 M |\qq| z-\zo} \left\{\log\left[\frac{\alm}{\alp} 
  \left(z-\frac{y_1}{M}\right) \left(z-\frac{y_2}{M}\right)\right]
  - \log\left[\left(z-\frac{z_1}{M}\right)\left(z-\frac{z_2}{M}\right)
\right]\right\} \nonumber\\
 & + & \frac{1}{(4\pi)^2} \int_{\alpha/M}^1 \!\! dz 
\frac{-1}{4 M |\qq| z-\zo} \left\{\log\left[ 
  \left(z-\frac{x_1}{M}\right) \left(z-\frac{x_2}{M}\right)\right]
  - \log\left[\left(z-\frac{z_1}{M}\right)\left(z-\frac{z_2}{M}\right)
\right]\right\}.
\end{eqnarray}
In order to be able to distribute the integrals inside the braces
without introducing spurious singularities at the zero of the
denominator, we add and subtract the value of each logarithm at $z =
\zo/(4 M |\qq|)$, with the help of identity (\ref{eq:ident}).  We can
then use the fact that $x_{1,2}$ have imaginary parts of opposite
sign, and analogously $y_{1,2}$ and $z_{1,2}$, to split the
logarithms.  (In the case of the logarithm containing $y_{1,2}$,
proper account must be taken of the fact that there is a factor
$\alm/\alp$ in the argument.  Whereas $\alp > 0$, we have $\alm > 0$
if $q^2 > 0$, and $\alm < 0$ if $q^2 < 0$.)  This procedure, which is
described in detail in \cite{thooft} (see also \cite{bouzas}, and
appendix \ref{sec:appa}), leads to a set of integrals that can be
evaluated in terms of dilogarithms, with the result,
\begin{eqnarray}
  \label{eq:fj3f}
  \J{3} & = & \lim_{M\rightarrow\infty} M \fJ{3} = \frac{1}{(4\pi)^2}
  \frac{1}{4 |\qq|} \sum_{j=1,2} 
  \left( \F{1} (y_j) + \F{2} (x_j) - \F{3} (z_j)  \right) \\
  \F{1} (x) & = & \Sp{\frac{\zo - 4 |\qq| \alpha}{\zo - 4 |\qq| x}} -
  \Sp{\frac{\zo}{\zo - 4 |\qq| x}} \nonumber \\
  \F{2} (x) & = & - \Sp{\frac{\zo - 4 |\qq| \alpha}{\zo - 4 |\qq| x}}
  - \frac{1}{2} \log^2 \left( \frac{\zo - 4 |\qq| x}{\mu^2}\right)
  \nonumber \\
  \F{3} (x) & \equiv  & \F{1} (x) + \F{2} (x)  = - \Sp{\frac{\zo}{\zo
  - 4 |\qq| x}} - \frac{1}{2} \log^2 \left( \frac{\zo - 4 |\qq|
  x}{\mu^2}\right). \nonumber
\end{eqnarray}
In this equation the functions $\F{2}$ and $\F{3}$ are defined in
terms of an arbitrary mass parameter $\mu>0$, analogous to the
dimensional regularization mass unit.  $\J{3}$ does not depend on
$\mu$ since, by virtue of the identity (\ref{eq:ident}), all $\mu$
dependence cancels in the sum in (\ref{eq:fj3f}).  In principle, $\mu$
can be eliminated from $\J{3}$, but we prefer not to do so.  In
appendix \ref{sec:appa} we discuss the form of $\F{2}$ in more detail,
and give several equivalent forms for it, some of which will be used
below. (\ref{eq:fj3f}) is then our final result for $\J{3}$.

The cuts of $\J{3}$ as a function of external momenta can be
easily found from (\ref{eq:fj3e}).  As $M\rightarrow\infty$, the
integral of the logarithm containing $y_{1,2}$ has a cut at
$(m^\prime+m)^2 < 4 q^2 < +\infty$, that of the log containing
$x_{1,2}$ has a cut at $\delta M/2 + m < v\cdot p<+\infty$, and the
integral of the log containing $z_{1,2}$ has a cut at $\delta M/2 +
m^\prime < v\cdot p^\prime<+\infty$.

\subsection{The massless case}
\label{sec:massless}

In HQET the heavy quark-gluon vertex at one loop \cite{grozin} is
expressed in terms of integrals of the form (\ref{j3}) with $m^\prime
= 0 = m$, and $\delta M = 0$ if the heavy quark residual mass term in
the Lagrangian is chosen to vanish.  The scalar integral $\J{3}$ has
been given in \cite{grozin} in that case, for $q^2 < 0$, $v\cdot
p^\prime < 0$, $v\cdot p < 0$.  It is then of interest to evaluate our
result (\ref{eq:fj3f}) in the massless case, both because of its
applicability to HQET and to cross-check our calculation with that of
\cite{grozin}.

Defining $w^\prime \equiv v\cdot p^\prime$, $w = v\cdot p$, $Q = 2
|\qq | = |\mathbf{p}^\prime -\mathbf{p}| > 0$,
$\sigma_{w^\prime}=\sgn(w^\prime)$ and $\sigma_{w}=\sgn(w)$, and
setting $m^\prime = m = \delta M = 0$, from (\ref{eq:fj3f}) we get,
\begin{eqnarray}
  \label{eq:massless}
\lefteqn{  \J{3}  =  \frac{1}{(4\pi)^2} \frac{1}{4|\qq|} \left\{ 
    \Sp{\frac{w^\prime+w-Q}{w^\prime + w + Q + i\varepsilon\sigma}}
    - \frac{1}{2} \log^2\left(\frac{1}{\mu^2} (w^\prime + w +
      Q) (w^\prime - w - Q) - i \varepsilon \sigma_w \right)
    \right.}\nonumber\\ 
    & - & \Sp{\frac{w^\prime+w+Q}{w^\prime + w - Q -
    i\varepsilon\sigma}} -\frac{1}{2} \log^2\left(\frac{1}{\mu^2}
    (w^\prime - w + Q) (w^\prime + w - Q) + i \varepsilon \sigma_w
    \right) \\
    & - & \Sp{\frac{(w^\prime-w+Q)(w^\prime+w-Q)}{(w^\prime - w - Q)
    (w^\prime+w+Q)- i\varepsilon\sigma_w}} + \frac{1}{2}
    \log^2\left(\frac{1}{\mu^2} (w^\prime + w - Q) (w^\prime - w - Q)
    - i \varepsilon \sigma_{w^\prime} \right) \nonumber\\
    & + & \left. \Sp{\frac{(w^\prime+w+Q)(w^\prime-w+Q)}{(w^\prime + w
    - Q)(w^\prime-w-Q)- i\varepsilon\sigma_{w^\prime}}}  
    +\frac{1}{2} \log^2\left(\frac{1}{\mu^2} (w^\prime + w +
      Q) (w^\prime - w + Q) + i \varepsilon \sigma_{w^\prime} \right)
    \right\} \nonumber.
\end{eqnarray}
This expression is in full numerical agreement with eq.\ (24) of
\cite{grozin} in the region where $q^2$, $w$ and $w^\prime$ are all
negative.  Also, (\ref{eq:massless}) has cuts at $w^\prime > 0$, $w>0$
and $q^2 > 0$, as it should.

\subsection{The limit $q\rightarrow 0$}
\label{sec:q=0}

In the limit $q\rightarrow 0$ the integral $\J{3}$ can be easily
computed as the difference of two-point integrals.  Using the notation
of \cite{bouzas} (see appendix \ref{sec:appb}), we have,
\begin{eqnarray}
  \label{eq:q=0}
  \J{3}|_{q=0} & = & \frac{1}{m^{\prime 2}-m^2}
  \left(\I{2}(\Delta,m^{\prime 2}) - \I{2}(\Delta,m^{2}) \right)
  \nonumber\\
  & = & \frac{1}{(4\pi)^2} \frac{1}{m^{\prime 2}-m^2} \left\{
  \frac{\Delta}{2} \log{\frac{m^2}{m^{\prime 2}}} +
  m^\prime {\cal F}(\xi^\prime) - m {\cal F}(\xi)  
  \right\}\\
  {\cal F}(x) & \equiv & \sqrt{x^2 - 1 + i\varepsilon} \left[
    \loga{x-\sqrt{x^2 - 1 + i\varepsilon}} -
    \loga{x+\sqrt{x^2 - 1 + i\varepsilon}}\right]\nonumber,
\end{eqnarray}
where $\xi^\prime=\Delta/(2 m^\prime)$, $\xi=\Delta/(2 m)$. 

In order to obtain the value at $q=0$ of $\J{3}$ directly from
(\ref{eq:fj3f}) we have to expand the functions in the sum to first
order in $q^\mu$, due to the factor $1/|\qq|$ in front of the sum. 

We consider first the terms $\sum_j (\F{1}(y_j)-\F{1}(z_j))$ in
(\ref{eq:fj3f}).  In the limit $q\rightarrow 0$ the two dilogarithms
in the definition of $\F{1}$ (see (\ref{eq:fj3f})) cancel each other
up to terms of $\oo (|\qq|^2)$, so these terms give a vanishing
contribution to $\J{3}$ at $q=0$.  It should be noted, however, that
the expansion about $q=0$ of $y_{1,2}$ is singular because of the
factor $1/\alpha_-$ in their definition (\ref{eq:roots}).  Thus, we
expand first about $|\qq|=0$, $q^0 > 0$ and take the limit $q^0
\rightarrow 0$ afterwards.  There are no such complications in the
case of $\F{1}(z_{1,2})$.

In order to expand the remaining terms $\sum_j
(\F{2}(x_j)-\F{2}(z_j))$ in (\ref{eq:fj3f}) about $q=0$, we use the
form (\ref{eq:a4}) for $\F{2}(x)$.  To lowest order, we get,
\begin{eqnarray}
  \label{eq:q=0a}
  \J{3} & = & \frac{1}{4|\qq|} \sum_{j=1,2} (\F{2}(x_j)-\F{2}(z_j)) +
  \oo (q) \nonumber\\
  & = & \sum_{j=1,2} \left\{
    - \frac{\overline x_j}{\overline \zo} + \frac{\overline x_j}{\overline \zo}
    \loga{-\frac{\overline x_j}{\mu}} +
    \frac{\overline z_j}{\overline \zo} - \frac{\overline z_j}{\overline \zo}
  \loga{-\frac{\overline z_j}{\mu}}
  \right\}.
\end{eqnarray}
Here, $\overline x_j$, $\overline z_j$, $\overline \zo$ are the
zeroth-order terms in the expansions of $x_j$, $z_j$, $\zo$ about
$q=0$, respectively.  Using the notation of (\ref{eq:q=0}), we have,
\begin{equation}
  \label{eq:roots0}
  \overline x_{1,2} = m \left(-\xi\pm \sqrt{\xi^2-1+i\varepsilon}\right),
  \quad
  \overline z_{1,2}  =  m^\prime \left(-\xi^\prime\pm \sqrt{\xi^{\prime
        2}-1+  i\varepsilon}\right),
  \quad
  \overline \zo =  -m^{\prime 2} + m^2 .
\end{equation}
Substituting (\ref{eq:roots0}) into (\ref{eq:q=0a}) we recover
(\ref{eq:q=0}).

\section{Tensor integrals}
\label{sec:tensorint}

In this section we consider tensor three-point integrals, up to the
third rank.  These are given in terms of integrals with smaller ranks
and fewer points, by using the well-known method of \cite{passar}.
The general tensor integral has the form (\ref{j3}).  We will restrict
ourselves to integrals of the standard form
\begin{equation} \label{t1}
  \JJ{3}^{\alpha_1\cdots \alpha_n} (v,q;\Delta,m,m^\prime)
  =  \frac{i\mu^{4-d}}{(2\pi)^d}
  \int\! d^d \ell \frac{\ell^{\alpha_1}\cdots \ell^{\alpha_n}}{
    \left(2v\!\cdot\! \ell-\Delta+i\varepsilon\right)
    \left((\ell+q)^2-m^2+i\varepsilon\right)
    \left((\ell-q)^2-m^{\prime 2}+i\varepsilon\right)}. 
\end{equation} 
By shifting $\ell\rightarrow \ell + (p^\prime +
p)/2$ in (\ref{j3}) we can express $\J{3}^{\alpha_1 \cdots \alpha_n}$
in terms of $\JJ{3}^{\alpha_1\cdots \alpha_n}$ as,
\begin{equation}
  \label{eq:t2}
  \J{3}^{\alpha_1 \cdots \alpha_n} = \JJ{3}^{\alpha_1\cdots \alpha_n}
  (v,q;\Delta,m,m^\prime) + \sum_{j=1}^n r^{\{\alpha_1} \cdots
  r^{\alpha_j} \JJ{3}^{\alpha_{j+1}\cdots \alpha_{n\}}}(v,q;\Delta,m,m^\prime),
\end{equation}
where $r^\mu\equiv 1/2 (p^\prime+p)^\mu$, and $A^{\{\alpha_1\alpha_2\cdots
  \alpha_s\}} \equiv A^{\alpha_1\alpha_2\cdots\alpha_s} +
  A^{\alpha_2\cdots\alpha_s\alpha_1} + \cdots +
  A^{\alpha_s\alpha_1\cdots\alpha_{s-1}}$.
Clearly, the scalar integral $\J{3}=\JJ{3}$.   The standard form
(\ref{t1}) has the advantage that it depends explicitly on only
two four-vectors, $v^\mu$ and $q^\mu$, instead of three as in
(\ref{j3}).  

\subsection{Vector integral}
\label{sec:vectorint}

For the vector integral we write $ \JJ{3}^\alpha (v,q; \Delta,m_1,m_2)
= \VV_1 v^\alpha + \VV_2 q^\alpha/|\qq|$, with $|\qq| \equiv
\sqrt{(v\cdot q)^2 -q^2}$, and $\VV_{1,2}=\VV_{1,2}(v\cdot
q,q^2,\Delta,m_1,m_2).$ If $|\qq|= 0$, then $q^\alpha\propto v^\alpha$
and we can set $\VV_2=0$.  If $|\qq|\neq 0$, then,
\begin{equation}
  \label{eq:vec2}
  |\qq|^2 \VV_1  = -q^2 v_\alpha \JJ{3}^\alpha + v\!\cdot\! q\, q_\alpha
   \JJ{3}^\alpha,
  \quad
  |\qq| \VV_2  =  v\!\cdot\! q\, v_\alpha \JJ{3}^\alpha - q_\alpha
  \JJ{3}^\alpha,
\end{equation}
with, 
\begin{eqnarray}
  v_\alpha \JJ{3}^\alpha(v,q;\Delta,m_1,m_2) & = & \frac{1}{2} B_0(4q^2,m_1,m_2) +
  \frac{\Delta}{2} \JJ{3}(v,q;\Delta,m_1,m_2) \nonumber\\
  q_\alpha \JJ{3}^\alpha(v,q;\Delta,m_1,m_2) & = & \frac{1}{4} \I{2}(\Delta-2
  v\cdot q, m_2) - \frac{1}{4} \I{2}(\Delta+2 v\cdot q, m_1)
    \label{eq:vec3}\\ 
  & & +\frac{m_1^2-m_2^2}{4} \JJ{3}(v,q; \Delta,m_1,m_2).\nonumber
\end{eqnarray}
The scalar two-point integrals $\I{2}$ and $B_0$ are given in the
appendix. 

\subsection{Second-rank tensor integral}
\label{sec:2tensor}

The decomposition of $\JJ{3}^{\alpha\beta}$ in terms of form factors
can be written as,
\begin{equation}
  \label{eq:tens1}
  \JJ{3}^{\alpha\beta} (v,q;\Delta,m_1,m_2)
  = \TT_1 g^{\alpha\beta} + \TT_2 v^\alpha v^\beta +
  \TT_3 \frac{q^\alpha q^\beta}{|\qq|^2} + \TT_4 \frac{v^{\{\alpha}
  q^{\beta\}}}{|\qq|}. 
\end{equation}
The components of $\JJ{3}^{\alpha\beta}$ are obtained by direct
computation,
\begin{eqnarray}
  \label{eq:tens3}
  \FF_1 & \equiv & g_{\alpha\beta} \JJ{3}^{\alpha\beta} = 
  \frac{1}{2} \I{2}(\Delta-2 v\cdot q,m_2) + \frac{1}{2}
  \I{2}(\Delta+2 v\cdot q,m_1) + \frac{m_1^2+m_2^2-2 q^2}{2}
  \JJ{3}\nonumber\\
  \FF_2 & \equiv & v_\alpha v_\beta \JJ{3}^{\alpha\beta} = -v\cdot q
  B_1\left(4 q^2,m_1,m_2\right) + \frac{1}{2}
  \left(\frac{\Delta}{2}-v\!\cdot\! q\right) 
  B_0\left(4 q^2,m_1,m_2\right) + \frac{\Delta^2}{4} 
  \JJ{3}\\
  \FF_3 & \equiv & q_\alpha q_\beta \JJ{3}^{\alpha\beta} =
  \frac{q_\alpha}{4} \left( \I{2}^\alpha(v,\Delta-2 v\cdot q,m_2) - 
    \I{2}^\alpha(v,\Delta+2 v\cdot q,m_1)\rule{0ex}{2ex}\right)
  \nonumber\\
  & & + \frac{q^2}{4} \left(\I{2}(\Delta-2 v\cdot q,m_2)+ \I{2}(\Delta+2
  v\cdot q,m_1) \rule{0ex}{2ex}\right) + \frac{m_1^2-m_2^2}{4}
  q_\alpha \JJ{3}^\alpha \nonumber\\
  \FF_4 & \equiv & v_\alpha q_\beta \JJ{3}^{\alpha\beta} = 
  -q^2 B_1(4 q^2,m_1,m_2)
  - \frac{q^2}{2} B_0(4 q^2,m_1,m_2) + \frac{\Delta}{2} q_\alpha
  \JJ{3}^\alpha 
  \nonumber 
\end{eqnarray}
where, on both sides of these equations the arguments
$(v,q;\Delta,m_1,m_2)$ of $\JJ{3}$, $\JJ{3}^\alpha$ and
$\JJ{3}^{\alpha\beta}$ have been omitted for brevity.  To one-loop
level, the form factors in (\ref{eq:tens1}) are given in terms of the
${\FF}s$ as,
\begin{eqnarray}
  \label{eq:tens3.5}
  2 |\qq|^2 \TT_1 & = & \left(1+\frac{\epsilon}{2}\right) \left( |\qq|^2
  \FF_1 + q^2 \FF_2 + \FF_3 - 2 v\!\cdot\! q \FF_4 \right)
  \nonumber \\ 
  2 |\qq|^2 \TT_2 & = & \left(1+\frac{\epsilon}{2}\right) q^2 \FF_1 +
  \left(3+\frac{\epsilon}{2}\right) \frac{q^4}{|\qq|^2} \FF_2 +
  \frac{1}{|\qq|^2} \left(2 (v\!\cdot\! q)^2 + 
    \left(1+\frac{\epsilon}{2}\right) q^2 \right) \FF_3 -
  (6+\epsilon) \frac{q^2 v\!\cdot\! q}{|\qq|^2} \FF_4
  \nonumber \\ 
  2 |\qq|^2 \TT_3 & = & \left(1+\frac{\epsilon}{2}\right) |\qq|^2 \FF_1
  + \left( 2 (v\!\cdot\! q)^2 + \left(1+\frac{\epsilon}{2}\right) q^2
  \right) 
  \FF_2 + \left( 3 + \frac{\epsilon}{2}  \right) \FF_3 - 
  (6 + \epsilon) v\!\cdot\! q \FF_4
  \\ 
  2 |\qq|^2 \TT_4 & = & - \left(1+\frac{\epsilon}{2}\right) v\!\cdot\! q
  |\qq|  \FF_1 - \left(3+\frac{\epsilon}{2}\right) \frac{q^2
  v\!\cdot\! q}{|\qq|} \FF_2 - \left(3+\frac{\epsilon}{2}\right)
  \frac{v\!\cdot\! q}{|\qq|} \FF_3 + \frac{1}{|\qq|} \left( 2q^2 + (4
  + \epsilon)
  (v\!\cdot\! q)^2 \right) \FF_4~.
  \nonumber  
\end{eqnarray}
Equations  (\ref{eq:tens3}) and (\ref{eq:tens3.5}) give an
explicit expression for $\JJ{3}^{\alpha\beta}.$

\subsection{Third-rank tensor integral}
\label{sec:3tensor}

The decomposition in terms of form factors can be written as,
\begin{eqnarray}
  \label{eq:tens4}
    \lefteqn{\JJ{3}^{\lambda\mu\nu} (v,q;\Delta,m_1,m_2)}
    \nonumber\\
  & = & \SS_1 g^{\{\lambda\mu}v^{\nu\}} + \SS_2 g^{\{\lambda\mu}q^{\nu\}}
  + \SS_3 v^{\{\lambda} v^\mu q^{\nu\}} + \SS_4 v^{\{\lambda} q^\mu
  q^{\nu\}} + \SS_5 v^{\lambda} v^\mu v^{\nu} + \SS_6 q^{\lambda} q^\mu
  q^{\nu} 
\end{eqnarray}
These form-factors are given implicitly by the following linear
relations,
\begin{eqnarray}
  (d+2) \SS_1+(d+2) v\!\cdot\! q \SS_2 + 3 v\!\cdot\! q \SS_3 + (q^2+2
  (v\!\cdot\! q)^2) \SS_4 + \SS_5 + q^2 v\!\cdot\! q  \SS_6 \nonumber & = &
  g_{\lambda\mu} v_\nu \JJ{3}^{\lambda\mu\nu}\nonumber\\
  (d+2) v\!\cdot\! q \SS_1 + (d+2) q^2 \SS_2 + (q^2+2 (v\!\cdot\!
  q)^2) \SS_3 + 3 q^2 v\!\cdot\! q  \SS_4 + v\!\cdot\! q \SS_5 + q^4 \SS_6
  & = & g_{\lambda\mu} q_\nu \JJ{3}^{\lambda\mu\nu} \nonumber\\
  3 v\!\cdot\! q \SS_1 + (q^2+2 (v\!\cdot\! q)^2) \SS_2 + (q^2+2
  (v\!\cdot\! q)^2) \SS_3 + (2 q^2+ (v\!\cdot\! q)^2) v\!\cdot\! q
  \SS_4 & &      \label{eq:tens5}\\
  + v\!\cdot\! q \SS_5 + q^2 v\!\cdot\! q \SS_6
  & = &   v_\lambda v_\mu q_\nu \JJ{3}^{\lambda\mu\nu}\nonumber\\
  (q^2+2 (v\!\cdot\! q)^2) \SS_1 +3 q^2 (v\!\cdot\! q) \SS_2 +
  v\!\cdot\! q ((v\!\cdot\! q)^2+2 q^2) \SS_3 + q^2 (q^2+ 2(v\!\cdot\!
  q)^2) \SS_4  & & \nonumber\\ 
  + (v\!\cdot q)^2 \SS_5+ q^2 v\!\cdot\! q \SS_6 
  & = & v_\lambda q_\mu q_\nu \JJ{3}^{\lambda\mu\nu}\nonumber\\
  3 \SS_1 + 3 v\!\cdot\! q\SS_2 + 3v\!\cdot\! q \SS_3 + 3 (v\!\cdot\!
  q)^2 \SS_4 + \SS_5+(v\!\cdot\! q)^3 \SS_6
  & = & v_\lambda v_\mu v_\nu \JJ{3}^{\lambda\mu\nu}\nonumber\\
  3 q^2 v\!\cdot\! q \SS_1 + 3 q^4 \SS_2 + 3 (v\!\cdot\! q)^2 q^2 \SS_3 + 3 (v\!\cdot\!
  q) q^4 \SS_4 + (v\!\cdot\! q)^3 \SS_5+ q^6 \SS_6
  & = & q_\lambda q_\mu q_\nu \JJ{3}^{\lambda\mu\nu}\nonumber,
\end{eqnarray}
where the right-hand side can be obtained by direct computation,
\begin{eqnarray}
  g_{\lambda\mu} v_\nu \JJ{3}^{\lambda\mu\nu} & = & \frac{1}{4} \left(
  A_0(m_1) + A_0(m_2) \right) + \frac{m_1^2+m_2^2-2 q^2}{4} B_0
  + \frac{\Delta}{2} \JJ{3\alpha}^\alpha \nonumber\\
  g_{\lambda\mu} q_\nu \JJ{3}^{\lambda\mu\nu} & = & \frac{q_\alpha}{2}
  \left(\I{2}^\alpha(v,\Delta+2 v\!\cdot\! q,m_1)  
   +
    \I{2}^\alpha(v,\Delta-2 v\!\cdot\! q,m_2)\right) \nonumber\\
  &  & + \frac{q^2}{2}
  \left(\I{2}(\Delta-2 v\!\cdot\! q,m_2)  -
    \I{2}(\Delta+2 v\!\cdot\! q,m_1)\right) 
  + \frac{m_1^2+m_2^2-2 q^2}{2} q_\alpha \JJ{3}^\alpha \nonumber\\
  v_\lambda v_\mu q_\nu \JJ{3}^{\lambda\mu\nu} & = &
  \frac{v_\alpha q_\beta}{2} B_2^{\alpha\beta}(2q,m_1,m_2) + 2 q^2
  v\!\cdot\! q B_1 + \frac{1}{2} v\!\cdot\! q q^2
  B_0 + \frac{\Delta}{2} v_\alpha q_\beta
  \JJ{3}^{\alpha\beta} \nonumber\\
  v_\lambda q_\mu q_\nu \JJ{3}^{\lambda\mu\nu} & = & \frac{1}{2}
  q_\alpha q_\beta B_2^{\alpha\beta}(2q,m_1,m_2) + \frac{q^4}{2} B_0
  + 2 q^4 B_1 + \frac{\Delta}{2}
  q_\alpha q_\beta \JJ{3}^{\alpha\beta}   \label{eq:tens6}\\
  v_\lambda v_\mu v_\nu \JJ{3}^{\lambda\mu\nu} & = &
  \frac{1}{2}v_\alpha v_\beta B_2^{\alpha\beta}(2q,m_1,m_2) +
  \frac{1}{2} (v\!\cdot\! q)^2 B_0 + 2 (v\!\cdot\! q)^2
  B_1 + \frac{\Delta}{2} v_\alpha v_\beta
  \JJ{3}^{\alpha\beta} \nonumber\\
  q_\lambda q_\mu q_\nu \JJ{3}^{\lambda\mu\nu} & = & \frac{1}{4}
  q_\alpha q_\beta \left(\I{2}^{\alpha\beta}(v,\Delta-2 v\!\cdot\!
    q,m_2) - \I{2}^{\alpha\beta}(v,\Delta+2 v\!\cdot\! q,m_1)\right)
  \nonumber\\
  & & +
  \frac{q^2}{2} q_\alpha \left(\I{2}^\alpha(v,\Delta-2 v\!\cdot\!
  q,m_2) + \I{2}^\alpha(v,\Delta+2 v\!\cdot\! q,m_1) \right) \nonumber\\
  & & 
  + \frac{q^4}{4} \left(\I{2}(\Delta-2 v\!\cdot\! q,m_2) -
  \I{2}(\Delta+2 v\!\cdot\! q,m_1)\right)
  + \frac{m_1^2-m_2^2}{4} q_\alpha q_\beta \JJ{3}^{\alpha\beta}.
  \nonumber
\end{eqnarray}
In these equations $B_0=B_0(4 q^2,m_1,m_2)$ and similarly
$B_1$, and $\JJ{3}=\JJ{3}(v,q;\Delta,m_1,m_2)$ and similarly
$\JJ{3}^\alpha$ and $\JJ{3}^{\alpha\beta}$.

\section{Final remarks}

In phenomenological applications, the exact functional dependence of
Feynman integrals on masses and residual momenta is usually not
needed.  Often, the first few terms in a series expansion in some of
the parameters provides the required accuracy.  We believe, however,
that the exact analytic computation presented here does not require
more calculational effort than approximate schemes.  It has the added
advantage of being valid over the entire physical region for internal
and external masses.  Applications of the results presented here to
loop graphs in heavy baryon chiral perturbation theory which have a
calculable dependence on the ratio of the $\Delta$-nucleon mass
difference to the pion mass will be given elsewhere.

Our approach, which is based on well-known methods and results for
vertex integrals in renormalizable theories \cite{thooft,passar},
streamlines the computation so that it can be easily reproduced and
verified, and adapted to other, more complicated one-loop diagrams.
Another important feature of this approach is that contact with the
unitary cuts of the diagram, as given by Cutkosky rules, is explicitly
kept at every step of the calculation \cite{thooft}.

\section*{Acknowledgements}

This work has been partially supported by Conacyt of Mexico.

\pagebreak
\appendix

\section{Calculation of $\F{2}$}
\label{sec:appa}
\setcounter{equation}{0}
\renewcommand{\theequation}{\thesection.\arabic{equation}}

In this appendix we discuss the calculation of the second integral in
(\ref{eq:fj3e}).  We consider first the basic integral, whose
computation is standard \cite{thooft}, 
\begin{eqnarray}
  \label{eq:a1}
  \lefteqn{\int_{\alpha/M}^1 \!\! dz 
  \frac{-1}{4 M |\qq| z-\zo} \left\{\loga{z-\frac{x}{M}}  -
  \loga{\frac{\zo}{4 |\qq| M}-\frac{x}{M}}\right\} 
   =  \frac{1}{4 M |\qq|} \left\{ \rule{0ex}{4ex}\frac{\pi^2}{6}
  +\Sp{\frac{4 |\qq|(\alpha-x)}{\zo-4 |\qq| x}}\right.} \nonumber\\ 
  &  &  \left. + \loga{\frac{\zo-4 |\qq| \alpha}{\zo-4 |\qq|x}}
  \loga{\frac{\alpha-x}{M}} - 
  \loga{\frac{\zo-4 |\qq| \alpha}{\zo-4|\qq|x}} 
  \loga{\frac{\zo-4 |\qq|x}{4 |\qq| M}}
   + \frac{1}{2} \log^2\left( \frac{-4 M |\qq|}{\zo-4 
   |\qq| x} \right)\right. \nonumber\\
  & & \left.  - \loga{\frac{-4 M |\qq|}{\zo-4|\qq|x}}
  \loga{\frac{4 M |\qq|}{\zo-4|\qq|x}} \rule{0ex}{4ex}\right\}  
  + \oo\left(\frac{1}{M^2}\right)
  \equiv \frac{1}{4 M |\qq|} \F{2}(x) +\oo\left(\frac{1}{M^2}\right),
\end{eqnarray}
where $\F{2}(x)$ is defined by this equation.  In (\ref{eq:a1}) we
have made use of the relation \cite{thooft} $\Sp{x} =
-\Sp{1/x}-\pi^2/6 -1/2 \log^2(-x)$ in order to obtain the leading $M$
dependence of $\mathrm{Li}_2 (4 M|\qq|/(\zo-4|\qq|x))$.

Denoting by $\mathcal{G}$ the second integral in (\ref{eq:fj3e}), we
have,
\begin{equation}
  \label{eq:a1.5}
  \mathcal{G} = \frac{1}{4 M |\qq|} \sum_{j=1,2}
  (\F{2}(x_j)-\F{2}(z_j)) + \oo\left(\frac{1}{M^2}\right).
\end{equation}
Clearly, those terms in $\F{2}(x)$ which do not depend on $x$ will cancel in
(\ref{eq:a1.5}), so they can be omitted.  Also, due to the identity
(\ref{eq:ident}),  $\mathcal{G}$ does not depend on the value of $M$,
which can be replaced by an arbitrary mass $\mu>0$ that remains constant
as $M\rightarrow\infty$.  Thus, in  (\ref{eq:a1.5}) we can write,
\begin{eqnarray}
  \label{eq:a2}
  \F{2}(x) & = & \Sp{\frac{4 |\qq|(\alpha-x)}{\zo-4
  |\qq| x}} + \loga{\frac{\zo-4 |\qq| \alpha}{\zo-4 |\qq|x}}
  \loga{\frac{\alpha-x}{\mu}} - 
  \loga{\frac{\zo-4 |\qq| \alpha}{\zo-4|\qq|x}} 
  \loga{\frac{\zo-4 |\qq|x}{4 |\qq| \mu}} \nonumber\\
  & & + \frac{1}{2} \log^2\left( \frac{-4 \mu |\qq|}{\zo-4
  |\qq| x} \right) - \loga{\frac{-4 \mu |\qq|}{\zo-4|\qq|x}}
  \loga{\frac{4 \mu |\qq|}{\zo-4|\qq|x}}. 
\end{eqnarray}
The last two logarithms in this equation can be rewritten according to
the identity $\log^2(-z)-2 \log(-z) \log(z) = -\pi^2 - \log^2(z)$,
which is valid for the principal determination of the log.  Dropping
the constant term, we see that in (\ref{eq:a1.5}) we can write,
\begin{eqnarray}
  \label{eq:a3}
  \F{2}(x) & = & \Sp{\frac{4 |\qq|(\alpha-x)}{\zo-4
  |\qq| x}} + \loga{\frac{\zo-4 |\qq| \alpha}{\zo-4 |\qq|x}}
  \loga{\frac{\alpha-x}{\mu}} - 
  \loga{\frac{\zo-4 |\qq| \alpha}{\zo-4|\qq|x}} 
  \loga{\frac{\zo-4 |\qq|x}{4 |\qq| \mu}}\nonumber \\
  & & - \frac{1}{2} \log^2\left( \frac{4 \mu |\qq|}{\zo-4
  |\qq| x} \right).
\end{eqnarray}
This equation can still be rewritten by completing the last two terms
to the square of a sum of logs.  Once again discarding $x$-independent
terms that do not contribute to (\ref{eq:a1.5}), we get,
\begin{equation}
  \label{eq:a4}
  \F{2}(x) = \Sp{\frac{4 |\qq|(\alpha-x)}{\zo-4
  |\qq| x}} + \loga{\frac{\zo-4 |\qq| \alpha}{\zo-4 |\qq|x}}
  \loga{\frac{\alpha-x}{\mu}} + \frac{1}{2} \log^2\left(
    \frac{\zo-4|\qq| x}{\zo-4|\qq| \alpha} \right).  
\end{equation}
From (\ref{eq:a3}), by means of the identity \cite{thooft}
\begin{eqnarray*}
  \label{eq:a5}
  \Sp{\frac{4 |\qq|(\alpha-x)}{\zo-4|\qq| x}} & = & 
  \Sp{1-\frac{\zo-4|\qq| \alpha}{\zo-4|\qq| x}} \\
  & = & \frac{\pi^2}{6} - \Sp{\frac{\zo-4|\qq| \alpha}{\zo-4|\qq| x}} 
  - \loga{\frac{\zo-4|\qq| \alpha}{\zo-4|\qq| x}}
  \loga{\frac{4 |\qq|(\alpha-x)}{\zo-4 |\qq| x}},
\end{eqnarray*}
we obtain $\F{2}(x)$ as given by (\ref{eq:fj3e}), up to constant
terms. All of these expressions for $\F{2}(x)$ lead to the same
results when substituted into (\ref{eq:a2}) or (\ref{eq:fj3e}).   

\pagebreak
\section{Loop integrals}
\label{sec:appb}

In this appendix we give a list of loop integrals used in the
foregoing.  More complete calculations can be found, e.g., in
\cite{bouzas,thooft,passar,denner,harlan} and references therein. 
Divergent integrals are separated in a dimensional-regularization pole
term and a finite remainder.  $\mub = \mu\sqrt{4\pi e^{-\gamma_E}}$.
\begin{eqnarray*}
  A_0(m) &=& \frac{i\mu^\epsilon}{(2\pi)^d} \int d^d\ell
  \frac{1}{\ell^2-m^2+i\varepsilon} = -\frac{m^2}{8\pi^2\epsilon} +
  \frac{m^2}{16\pi^2} a_0(m^2)\\
  a_0(m^2) &=& \log\left(\frac{m^2}{\mub^2}\right) - 1\\
%
%
%
%
  B_0(p^2,m_1,m_2) &=& \frac{i\mu^\epsilon}{(2\pi)^d} \int d^d\ell
  \frac{1}{\left(\ell^2-m_1^2+i\varepsilon\right)
    \left((\ell+p)^2-m_2^2+i \varepsilon\right)}\\
  &=& -\frac{1}{8\pi^2\epsilon} + \frac{1}{16\pi^2}
  b_0(p^2,m_1^2,m_2^2)\\
  b_0(p^2,m_1^2,m_2^2) &=& \int_0^1 dx \log\left((1-x)
  \frac{m_1^2}{\mub^2} + x \frac{m_2^2}{\mub^2} - x(1-x)
  \frac{p^2}{\mub^2} -i\varepsilon\right)\\
  B_1^\mu(p,m_1,m_2) &=& \frac{i\mu^\epsilon}{(2\pi)^d} \int
  d^d\ell 
  \frac{\ell^\mu}{\left(\ell^2-m_1^2+i\varepsilon\right)
    \left((\ell+p)^2-m_2^2+i \varepsilon\right)}
  = p^\mu B_1(p^2,m_1,m_2)\\
  p^2 B_1(p^2,m_1,m_2) &=& \frac{1}{2}\left(A_0(m_1) - A_0(m_2)
    - (p^2+m_1^2-m_2^2) B_0(p^2,m_1,m_2)\right)\\
  B_2^{\mu\nu}(p,m_1,m_2) &=& \frac{i\mu^\epsilon}{(2\pi)^d} \int d^d\ell
  \frac{\ell^\mu\ell^\nu}{\left(\ell^2-m_1^2+i\varepsilon\right)
    \left((\ell+p)^2-m_2^2+i \varepsilon\right)}
  = G \left(g^{\mu\nu}-\frac{p^\mu p^\nu}{p^2}\right) + L \frac{p^\mu
  p^\nu}{p^2}\\
  (d-1) G &=& \frac{1}{2} A_0(m_2) + m_1^2 B_0(p^2,m_1,m_2) +
  \frac{1}{2} (m_1^2-m_2^2+p^2) B_1(p^2,m_1,m_2)\\
  L &=& \frac{1}{2} A_0(m_2) -  \frac{1}{2} (m_1^2-m_2^2+p^2)
  B_1(p^2,m_1,m_2) \\
  \I{2}(\Delta,m) & = & \frac{i\mu^\epsilon}{(2\pi)^d} \int d^d\ell
  \frac{1}{(2v\!\cdot\!\ell-\Delta+i\varepsilon)(\ell^2-m^2+i\varepsilon)}\\
  & = & \frac{\Delta}{32\pi^2}\left(\frac{2}{\epsilon} +
  \loga{\frac{\mub^2}{m^2}} + 2 \right)  + \frac{m}{16\pi^2}
  \mathcal{F}\left(\frac{\Delta}{2m}\right)\quad \mbox{(see (\ref{eq:q=0}))} \\
  \I{2}^\mu(v,\Delta,m) & = & \frac{i\mu^\epsilon}{(2\pi)^d} \int d^d\ell
  \frac{\ell^\mu}{(2v\!\cdot\!\ell-\Delta+i\varepsilon)(\ell^2-m^2+i\varepsilon)} 
   =  F(\Delta,m) v^\mu\\
  F(\Delta,m) & = & \frac{1}{2} A_0(m) + \frac{\Delta}{2} \I{2}(\Delta,m)\\
  \I{2}^{\mu\nu}(v,\Delta,m) & = & \frac{i\mu^\epsilon}{(2\pi)^d} \int d^d\ell
  \frac{\ell^\mu\ell^\nu}{(2v\!\cdot\!\ell-\Delta+i\varepsilon)
   (\ell^2-m^2+i\varepsilon)} 
 =  I_0(\Delta,m) g^{\mu\nu} + I_1(\Delta,m) v^\mu v^\nu\\
 (d-1) I_0(\Delta,m) & = & -\frac{\Delta}{4} A_0(m) +
 \left(m^2-\frac{\Delta^2}{4}\right) \I{2}(\Delta,m) \\
 (d-1) I_1(\Delta,m) & = & \frac{d\Delta}{4} A_0(m) -
 \left(m^2-\frac{d\Delta^2}{4}\right) \I{2}(\Delta,m) 
\end{eqnarray*}
\end{document}